\documentstyle[psfig]{l-aa}              
\def\uz{UZ For}
\def\pe{$\phi_{\rm ecl}$}
\def\kmps{km\,s$^{-1}$}
\def\msun{M$_{\sun} $}
\def\mwd{$M_{\rm wd}$}
\begin{document}
\thesaurus {06(08.02.1, 08.02.2, 08.14.2, 08.09.2 UZ For, 02.01.2)}
\title{
On the mass of the white dwarf in UZ Fornacis\thanks{Based 
on observations collected at the European Southern
Observatory, La Silla, Chile.}}
  \author {Axel D.~Schwope\inst{1} 
	\and 
           Sabine Mengel\inst{1} 
	\and
	   Klaus Beuermann\inst{2}
	}
   \offprints{Axel D.~Schwope}
                               
  \institute {
	Astrophysikalisches Institut Potsdam, An der Sternwarte 16,
	D--14482 Potsdam, FRG {\it (e-mail: ASchwope@aip.de)}
\and	Universit\"atssternwarte G\"ottingen, Geismarlandstra\ss e 11, 
        D--37083 G\"ottingen, FRG
}
\date{Received, accepted}
\maketitle
\markboth{Schwope, A.D. et al.: Mass determination of UZ Fornacis}{}
\begin{abstract}
We present phase-resolved spectroscopy of the eclipsing AM Herculis star
UZ For obtained when the system was in its low state of accretion. 
Faint residual H$\alpha$-emission and NaI absorption were used to trace 
the secondary star and infer its orbital velocity $K_2$. The 
measured radial velocity amplitude of NaI $K_2 * \sin{i} = 
285 \pm 50$\,\kmps\ suggests a low-mass white dwarf with $M_{\rm wd} = 
0.44 \pm 0.15$\,\msun (1$\sigma$-errors).
The H$\alpha$ emission line on the other hand,
visible only for part of the orbital cycle and 
supposed to originate only on the illuminated hemisphere
facing the white dwarf, 
displays a similar radial velocity amplitude, $K'_2 \sin i = 
308\pm 27$\,\kmps. The standard $K_2$-correction applied by us then suggests
a white dwarf mass of up to 1\,\msun. Compared with earlier results
the new ones enlarge the
window in which the white dwarf mass may lie and resolves the 
conflict between mass estimates based on photometry and 
spectroscopy. They leave some ambiguity in the location of emission 
and absorption components in these and former observations.
\end{abstract}
\keywords{Accretion -- cataclysmic variables -- AM Herculis binaries -- 
             stars: \uz\ -- stars: eclipsing}

\section{Introduction}
The eclipsing polar (AM Herculis binary) UZ For became of considerable interest
soon after its discovery and optical identification (Osborne et al.~1988,
Beuermann et al.~1988, Paper I) because of its high magnetic field, its
period of 126.5\,min, close to the lower edge of the CV period gap, and
the likely high mass of the accreting white dwarf, $M_{\rm wd} > 0.93$\,\msun\/
(90\% confidence). The high mass of the white dwarf was based on the 
observed high radial velocity amplitudes of H$\alpha$ emission and NaI
absorption lines originating at the secondary star. 
The white dwarf in \uz\ thus seemed to be more massive than those
in other AM Her binaries and than single ones.
On the assumption that \uz\ resumes accretion at the 
observed period after crossing the period gap (a quantity which is dependent
on \mwd), Hameury et al.~(1988) argued also for a massive white dwarf from 
evolutionary considerations, \mwd\ $\simeq 1.2$\,\msun. 
High-speed photometry presented by
Bailey and Cropper (1991) with resolved ingress and egress of the white 
dwarf, however, suggests that the white dwarf is a normal one with
standard mass \mwd = 0.61 -- 0.79 \msun.

\section{Observations and analysis}
In order the resolve the conflict between the different mass estimates,
we re-observed \uz\ with the ESO 3.6m telescope equipped with EFOSC1 and
RCA CCD in the night Nov.~21/22, 1989. EFOSC1 allows 
low-resolution grism spectroscopy. The grism used by us, R300, yielded 
a FWHM spectral resolution of $\sim$25\AA\ and wavelength coverage 
5800 -- 10000 \AA. We obtained a total of 15 spectra with 
integration times of 480\,sec thus achieving full phase coverage.

Flux-calibration of the extracted and wavelength-calibrated spectra
was achieved by observations of the spectrophotometric standard star
Feige 110, whose spectrum was also used in order to derive a correction
function for absorption in the earth atmosphere. 
\uz\ was in a similar state of activity compared with our initial observations
in 1987 discussed in Paper I and by Schwope et al.~(1990). The reader is 
referred to these papers for mean orbital and typical bright- and 
faint-phase spectra. 

As in these former observations, faint H$\alpha$ emission
is present with strong photometric variability, being brightest at \pe
$\simeq 0.5$ and non-detectable around eclipse which suggests an origin on or
near the illuminated hemisphere. The positions of the H$\alpha$-lines in the 
individual spectra were measured by Gaussian fits. This was possible 
only in spectra covering the phase interval \pe = 0.36 -- 0.80 due to the
faintness of the line otherwise. A sine fit with fixed period, $P_{\rm orb}
= 7591.573$\,sec, and fixed phase zero at mid-eclipse yielded a radial
velocity amplitude $K_{\rm H\alpha} = K'_2 \sin i = 308 \pm 27$\,\kmps.

\begin{figure}[t]
\psfig{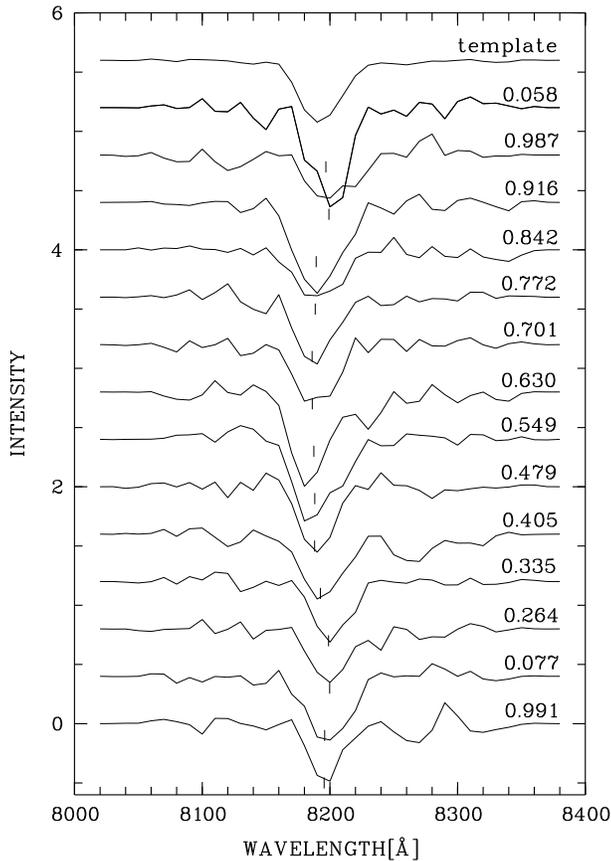}
\caption[meanlc]{\label{spec}
Normalized spectra of \uz\ centered on the NaI doublet at 8200\,\AA. The
uppermost curve is the template spectrum used for the crosscorrelation 
analysis. Small vertical ticks in the absorption lines
indicate the positions obtained by crosscorrelation. Units along the 
ordinate are $10^{-16}$\,erg cm$^{-2}$\,s$^{-1}$\,\AA$^{-1}$. The spectra are
shifted vertically by 0.4 units with respect to each other starting with the 
first in the sequence at bottom.
Numbers above the spectra indicate orbital phases.
}
\end{figure}

Due to the overall faintness of the system at the time of observation, 
photospheric radiation 
from the secondary star dominates at long wavelengths. The distorted shape
of the companion star gives rise to a pronounced ellipsoidal variation 
in the infrared. 
Photospheric NaI absorption from the secondary star is clearly present in
each of our spectra, although the doublet is spectrally not resolved. 
The spectra of \uz\ were rectified to continuum intensity zero by a two-stage
procedure. First, suitably scaled template spectra of the M dwarf Gl~83.1
with removed NaI-line were subtracted from the original spectra. Second,
low-order polynomials were fitted to the now (after step 1) smooth continuum
and subtracted. This yields spectra with continuum intensity zero 
and the NaI absorption line as the only prominent
structure. These are shown in original time sequence in Fig.~\ref{spec}. 

\begin{figure}[ht]
\psfig{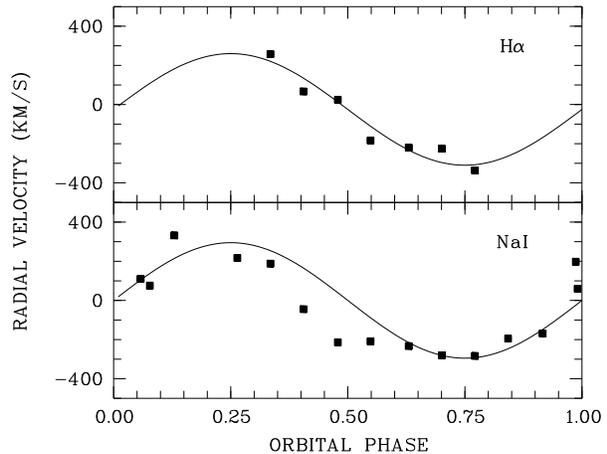}
\caption[vrad]{\label{vrad}
Radial velocity curves and best-fit sine curves for the H$\alpha$ emission and
NaI absorption lines in UZ For.
}
\end{figure}

We have measured the positions of individual lines by fitting Gaussian
absorption lines to the observed troughs. 
These positions were then fitted with a sine curve with fixed period
and phase zero, which yields a radial velocity amplitude 
$K_{\rm Na} = K_2 \sin i = 285 \pm 50$\,\kmps. 
We have checked these results by a
crosscorrelation analysis which yielded essentially the same results.
The measured radial velocities of the H$\alpha$- and NaI-lines are
shown together with the best-fit sine curves in Fig.~\ref{vrad}.

\section{Results and discussion}

Both radial velocity measurements, H$\alpha$ and NaI, can be used to 
re-estimate the 
mass of the white dwarf. This is done using the following assumptions:
(a) The secondary fills its Roche lobe completely and is (b) 
a main sequence star; (c) The mass-radius relation of the secondary 
is of the form 
$R_2 = \alpha M_2^\gamma$, and $R_2$ is set equal to the spherical 
equivalent Roche radius $R_s$; (d) $R_s$ is approximated by the formula
given by Eggleton (1983) as a function of the mass ratio $Q$;
(e) The measured radial velocity amplitude 
$K_{\rm Na}$ represents the true orbital velocity of the companion star
(apart from a $\sin{i}$ factor); (f) The full eclipse length of the white 
dwarf is 463.9\,sec corresponding to a half phase angle of 11.00\degr. 
Allen et al.~(1989) derived a full eclipse length of the hot accretion 
spot on the white dwarf surface of $466.5 \pm 2.5$\,sec. This number has 
to be corrected for the offset between the white dwarf center and the 
spot. We assume for this correction a binary separation of $a = 5.30 \times
10^{10}$\,cm, a white-dwarf radius $R = 8.7 \times 10^8$\,cm, an azimuth
of the spot of -45\degr\ and a colatitude of 150\degr.

\begin{figure}
\psfig{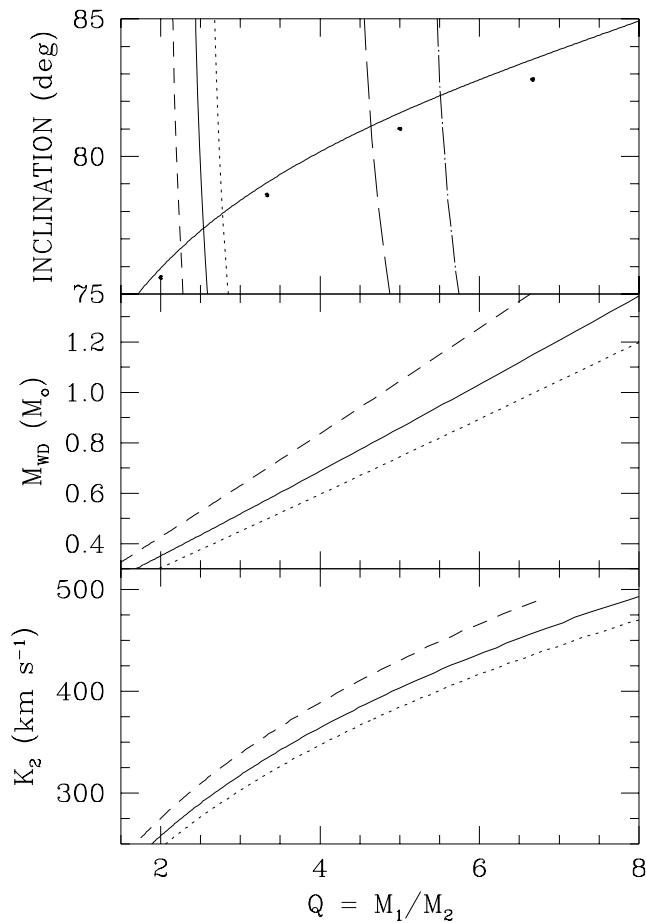}
\caption[meanlc]{\label{qirel}
Mass estimate of the white dwarf in \uz. The three panels show from top
to bottom the inclination $i$, the mass of the white dwarf
and the predicted orbital velocity of the secondary star 
as a function of the mass ratio $Q$ for 
given period, eclipse length and adopted mass-radius relation for the 
secondary star, respectively. In each of the panels the three more or 
less parallel 
lines indicate predictions made by different mass-radius relations
for the secondary star (solid:
Neece 1984, dotted: Caillault \& Patterson (1990), short dashed: VandenBerg 
et al.~1983). 
The solid line connecting the lower left with the upper right in the upper
panel relates the eclipse length with $i$ and $Q$ and was calculated for
an adopted eclipse length of 463.9 sec. The dots below this curve are
drawn from Bailey \& Cropper (1991) who used an unknown eclipse length.
The three mainly vertical lines to the left in the upper panel are mass
functions calculated for a velocity of 285\,\kmps\/and different M/R-relation.
The long-dashed line and the dashed-dotted line are mass functions for
385\,\kmps\/and 416\,\kmps, respectively, using the M/R-relation given 
by Neece.
}
\end{figure}

The relation between inclination $i$ and mass ratio $Q$ for given eclipse 
length due to Chanan et al.~(1976) is shown as a solid curve in the upper
panel of Fig.~\ref{qirel} connecting the lower left with the upper right edge. 
This relation is purely geometrical and not dependent
on e.g.~ZAMS mass-radius assumptions. It seems, however, that we used
a different eclipse length of the white dwarf than Bailey \& Cropper (1991)
did in their analysis (small dots below the solid curve are depicted from 
their Table 2). Unfortunately they do not quote their measured
eclipse length, so that a direct comparison is not possible. The highest
inclination and highest mass ratio compatible with the observed eclipse length
are $i_{\rm max} = 86.2\degr$ and $Q_{\rm max} = 9.20$. 

The three 
parallel lines in the upper panel to the left are the mass functions 
calculated for the nominal velocity of 285\,\kmps using three different
mass radius relations. The middle line is based on Neece (1984, N84),
the dotted line on Caillault \& Patterson (1990, CP90) 
and the short dashed line on VandenBerg et al.~(1983, VdB83). 
The range of predicted secondary star masses
by these authors is rather large, ranging from 0.149\,\msun (CP90)
and 0.174\,\msun (N84) 
to 0.212\,\msun (VdB83). The intersection points between
both ($Q,i$)-relations shown in the upper panel have to be reflected at
the corresponding lines in the second panel in order to read white-dwarf
mass for the nominal orbital velocity. This yields a rather low white-dwarf
mass, \mwd $\simeq 0.42 - 0.48$\,\msun. The high-mass limit is subject 
to the adopted mass-radius relation of the secondary star and the confidence
level one would like to reach. For the reader's convenience we 
show the three by the different $M/R$-relations 
predicted orbital velocity amplitudes for the secondary star as a function 
of $Q$ in the lower panel of Fig.~\ref{qirel}.
For 1, 2, 3 $\sigma$-errors of the radial 
velocity amplitude (335, 385, 435\,\kmps) and the mass-radius relation N84
one obtains $M_{\rm wd, max} = 0.6, 0.80, 1.04$\,\msun, respectively. As an 
illustration we show with long dashes the corresponding mass function for
the 2$\sigma$-level in the upper panel of Fig.~\ref{qirel}. 
The new NaI measurements
suggest a white-dwarf mass not in excess of 1\,\msun\ and a 
mass ratio not in excess of 6. 

Similarly the H$\alpha$-line can be used to estimate the white dwarf mass.
Its photometric and radial velocity variation suggest an origin on that 
hemisphere of the companion star which is illuminated by EUV-radiation from 
the accretion spot. The measured velocity amplitude has to be multiplied 
by a $Q$-dependent factor in order to scale the observed center-of-light 
velocity to the required center-of-mass velocity ($K_2$-correction). It 
is clear that the corresponding mass estimate must yield a higher value
than that using the NaI lines, since both measured velocity 
amplitudes are more or less the same.
A graph of the factor mentioned 
is shown in Schwope et al.~(1993), for $Q= 5.5$ its value
is 1.35. The corresponding $(Q,i)$-relation is shown 
with a dash-dotted line in Fig.~\ref{qirel} 
(upper panel) and suggests, as expected, a much 
higher mass of the white dwarf \mwd $ \simeq 0.95$\,\msun\ ($Q \simeq 5.5$)
than obtained using the NaI line.

The model for the $K_2$-correction assumes that reprocessed emission originates
from the whole illuminated part of the secondaries Roche lobe and that its 
intensity at given velocity (surface element) is proportional to the 
locally incident radiation. The real situation might be different and the 
mass estimate therefore in error. We have found recently two pronounced
examples for such a deviation in other AM Her systems. First, the 
H$\alpha$ emission line radial velocity in MR Ser seen in a low state
is the same as that of the NaI absorption line (as here in UZ For). This
velocity amplitude is much higher than that of the narrow emission line 
from the secondary star seen in MR Ser's high state (Schwope et al.~1993).
Second, Doppler tomography of the Balmer and HeII emission lines of 
HU Aqr observed in the high accretion state show that the Balmer emission 
is concentrated away from the inner Lagrangian point 
(Schwope et al.~1996). Both examples suggest that the reason for the 
negligible difference between the NaI and H$\alpha$ velocity amplitudes 
in the present case of \uz\
is a similar deviation of the Balmer emission from the simple pattern and that,
hence, the $K_2$-correction applied might be too large.
A common solution at the $\sim$2$\sigma$-level 
for both our radial velocity measurements is $Q=4.4$, \mwd $ = 0.75$, and 
$i = 80.7\degr$, which is also well compatible with the photometric 
study by Bailey \& Cropper (1991).

Finally we comment on the possible origin of the higher radial velocities
derived in Paper I. These data were obtained in a similar
state of accretion and with comparable time and spectral resolution, 
but with a  smaller telescope and with a high read-noise RCA CCD (as the
data presented here). This
allowed measurement of the NaI-lines in only a few spectra. It is possible
that these measurements were misleading. The number of useful H$\alpha$ 
measurements in both observations is the same, and the radial velocities
are not too discrepant ($395\pm67$\,\kmps at 90\% confidence in Paper I,
$308 \pm 27$\,\kmps (1$\sigma$) here). Since the H$\alpha$-line 
measures the center of light on the illuminated hemisphere of the secondary
with possible contaminations from the accretion  stream a discrepant 
radial velocity amplitude would simply mean, that the center of 
light is shifted between the different observations. The degree of freedom
for shifts of the NaI radial velocity, however, is smaller, 
although illumination might shift the center of light towards the 
non-illuminated hemisphere (e.g.~Davey \& Smith 1992).
 
Our main results are summarized: (1) The mass of the white dwarf in \uz\ is
most likely not in excess of 1\msun\ ($i < 83\degr, Q < 6$). (2) The 
mass-radius relation of low-mass main sequence stars introduces large
uncertainties to mass estimates of white dwarfs in polars and needs proper
calibration. (3) The ultimate location of the low-state Balmer emission on the 
illuminated secondary in \uz\ remains uncertain, hence the applicability 
of the $K_2$-correction scheme is questionable. One needs high signal-to-noise
data with high phase resolution in order to calibrate empirically
the $K_2$-correction. (4) The likely low mass of the white dwarf in UZ For 
together with its measured orbital period suggest that the system was born 
in or slightly above the period gap.

\acknowledgements
We thank an anonymous referee for helpful comments.
This work was supported by the BMFT under grant 50 OR 9403 5.

\end{document}